\documentclass[amsmath,aps,showpacs,a4paper,10pt]{revtex4}

 \usepackage{epsf}
 \usepackage{graphicx}    

\begin{document}

 \newcommand{\be}[1]{\begin{equation}\label{#1}}
 \newcommand{\ee}{\end{equation}}
 \newcommand{\bea}{\begin{eqnarray}}
 \newcommand{\eea}{\end{eqnarray}}
 \def\disp{\displaystyle}

 \def\gsim{ \lower .75ex \hbox{$\sim$} \llap{\raise .27ex \hbox{$>$}} }
 \def\lsim{ \lower .75ex \hbox{$\sim$} \llap{\raise .27ex \hbox{$<$}} }

 \begin{titlepage}

 \begin{flushright}
 arXiv:1310.5338
 \end{flushright}

 \title{\Large \bf Cosmological Evolution of Einstein-Aether Models
 with~Power-law-like Potential}

 \author{Hao~Wei\,}
 \email[\,email address:\ ]{haowei@bit.edu.cn}
 \affiliation{School of Physics, Beijing Institute
 of Technology, Beijing 100081, China}

 \author{Xiao-Peng~Yan\,}
 \affiliation{School of Physics, Beijing Institute
 of Technology, Beijing 100081, China}

 \author{Ya-Nan~Zhou\,}
 \affiliation{School of Physics, Beijing Institute
 of Technology, Beijing 100081, China}

 \begin{abstract}\vspace{1cm}
 \centerline{\bf ABSTRACT}\vspace{2mm}
 The so-called Einstein-Aether theory is General Relativity
 coupled (at second derivative order) to a dynamical unit
 time-like vector field~(the aether). It is a Lorentz-violating
 theory, and gained much attention in the recent years. In
 the present work, we study the cosmological evolution of
 Einstein-Aether models with power-law-like potential, by using
 the method of dynamical system. In the case without matter,
 there are two attractors which correspond to an inflationary
 universe in the early epoch, or a de~Sitter universe in the
 late time. In the case with matter but there is no interaction
 between dark energy and matter, there are only two de~Sitter
 attractors, and no scaling attractor exists. So, it is
 difficult to alleviate the cosmological coincidence problem.
 Therefore, we then allow the interaction between dark energy
 and matter. In this case, several scaling attractors can exist
 under some complicated conditions, and hence the cosmological
 coincidence problem could be alleviated.
 \end{abstract}


 \maketitle

 \end{titlepage}

 \renewcommand{\baselinestretch}{1.1}


\section{Introduction}\label{sec1}

The so-called Einstein-Aether~({\AE}) theory~\cite{r1,r2,r3,r4,r5,r6}
 is General Relativity (GR) coupled (at second derivative order)
 to a dynamical unit time-like vector field~(the aether). In
 this theory, the local structure of spacetime is described
 by the metric tensor $g_{ab}$ (as in GR) together with the
 aether vector $u_a$. The aether defines a preferred rest frame
 at each point of spacetime, but preserves rotational symmetry
 in that frame. The {\AE} theory is a Lorentz-violating theory,
 while it preserves locality, general covariance and the
 successes of GR~\cite{r1,r2,r3,r4,r5,r6}.

The {\AE} theory gained much attention in the recent years.
 For instance, some observational and experimental constraints
 on the {\AE} theory have been considered~\cite{r7,r8,r9,r36}
 (see also~\cite{r1}). It has been claimed in~\cite{r3} that
 if the aether is restricted in the action to be hypersurface
 orthogonal, the {\AE} theory is identical to the IR limit of
 the healthy extension of Horava-Lifshitz gravity~\cite{r10}.
 Black holes, neutron stars and binary pulsars in the {\AE}
 theory have been extensively considered in
 e.g.~\cite{r11,r12,r13,r14,r15,r36}. Also,
 supersymmetric aether has been discussed in
 e.g.~\cite{r16,r17}. In~\cite{r18}, magnetic fields in the
 {\AE} theory have been studied. In~\cite{r19}, Godel type
 metrics in the {\AE} theory were considered. The {\AE}
 theory as a quantum effective field theory was considered
 in~\cite{r21}. A positive energy theorem for the {\AE} theory
 has been shown in~\cite{r22}.

Of course, it is not surprising to find that the {\AE} theory
 has been extensively considered in cosmology. For example,
 some exact solutions of the inflationary {\AE} cosmological
 models have been found in~\cite{r5}. In~\cite{r23}, it has
 been considered as an alternative to dark matter or dark
 energy. Also, the cosmological perturbations in the {\AE}
 theory have been studied in~\cite{r24,r25}. The stability of
 {\AE} cosmological models has been discussed in~\cite{r6}. The
 inflaton coupled with an expanding aether was considered
 in~\cite{r2}. The cosmic microwave background polarization in
 the {\AE} theory has been investigated in e.g.~\cite{r26}.
 The observational constraints on the {\AE} models as an
 alternative to dark energy were obtained in e.g.~\cite{r27}.
 The cosmological constraints on modified Chaplygin gas in
 the {\AE} theory has also been found in~\cite{r28}. The
 anisotropic {\AE} cosmological models were
 discussed in~\cite{r29}. The Lorentz-violating inflation in
 the {\AE} theory has been studied in~\cite{r30}. In fact,
 there are many other interesting works on the {\AE} cosmology
 in the literature, and we cannot mention all of them here.

In a Friedmann-Robertson-Walker (FRW) universe, the aether
 field will be aligned with the cosmic frame, and is related
 to the expansion of the universe~\cite{r2,r5,r6}, namely
 \be{eq1}
 \nabla_c u_b=\frac{\theta}{3}\left(g_{cb}-u_c u_b\right)\,,
 \ee
 where the expansion $\theta=3H$ in the FRW cosmology, and
 $H\equiv\dot{a}/a$ is the Hubble parameter; $a$ is the
 scale factor; a dot denotes the derivative with respect
 to the cosmic time $t$. If there exists a self-interacting
 scalar field $\phi$ (which might play the role of inflaton
 or dark energy) in the universe, it will couple to the
 expanding aether. The modified stress tensor is given
 by~\cite{r2,r5,r6}
 \be{eq2}
 T_{ab}=\nabla_a \phi \nabla_b \phi-\left(
 \frac{1}{2}\nabla_c \phi \nabla^c \phi-V+\theta V_\theta
 \right)g_{ab}+\dot{V}_\theta\left(u_a u_b-g_{ab}\right)\,,
 \ee
 where the potential $V=V(\theta, \phi)$ is now a function of
 both $\theta$ and $\phi$\,. Notice that
 $V_\theta\equiv\partial V/\partial\theta$. So,
 the corresponding energy density $\rho_{\phi}$ and pressure
 $p_{\phi}$ of the homogeneous scalar field
 read~\cite{r2,r5,r6}
 \bea
 &&\rho_{\phi}=\frac{1}{2}\dot{\phi}^2+V-
 \theta V_\theta\,,\label{eq3}\\
 &&p_{\phi}=\frac{1}{2}\dot{\phi}^2-V
 +\theta V_\theta+\dot{V}_\theta\,.\label{eq4}
 \eea
 Using Eqs.~(\ref{eq3}) and (\ref{eq4}), it is easy to see
 that the energy conservation equation of the scalar field,
 \be{eq5}
 \dot{\rho}_\phi+\theta\left(\rho_\phi+p_\phi\right)=0\,,
 \ee
 is actually equivalent to the equation of motion
 \be{eq6}
 \ddot{\phi}+\theta\dot{\phi}+V_\phi=0\,,
 \ee
 where $V_\phi\equiv\partial V/\partial\phi$\,. Interestingly,
 in the {\AE} cosmology, Eq.~(\ref{eq6}) still has the same
 form as in GR.

In~\cite{r5}, Barrow proposed an exponential-like potential
 \be{eq7}
 V(\theta, \phi)=V_0\,e^{-\lambda\phi}+\sum\limits_{r=0}^n
 \mu_r \theta^r e^{(r-2)\lambda\phi/2}\,,
 \ee
 and found some exact solutions for the {\AE} models with
 this potential. Barrow~\cite{r5} noted that this choice of
 potential subsumes the simple cases with
 $V(\theta, \phi)=f(\phi)\theta^2$ of~\cite{r30} and
 $V(\theta, \phi)=f(\theta^2)$ of~\cite{r23}, but
 {\em cannot} include the choice
 $V(\theta, \phi)=\frac{1}{2}m^2 \phi^2+\mu\theta\phi$
 considered in~\cite{r2}. Then, Sandin {\it et al.}~\cite{r6}
 studied the cosmological evolution of the {\AE} models with
 a special exponential-like potential,
 \be{eq8}
 V(\theta, \phi)=V_0\,e^{-\lambda\phi}+
 \mu_1\sqrt{V_0}\,\theta\,e^{-\lambda\phi/2}+\mu_2 \theta^2\,,
 \ee
 and found some interesting results.

In the present work, we would like to suggest a power-law-like
 potential for the {\AE} models, namely
 \be{eq9}
 V(\theta, \phi)=V_0\,\phi^n+\sum\limits_{r=0}^m
 \mu_r \theta^r \phi^{n(2-r)/2}\,,
 \ee
 where $V_0$, $n$ and $\{\mu_r\}$ are all constants. There
 are two fine motivations to consider this type of potential.
 First, as is well known, the power-law potential is important
 and hence has been extensively used in the literature since
 it can rise from various fundamental theories. For instance,
 a power-law potential might be motivated from the dilatation
 symmetry~\cite{r20}. Although we cannot directly derive the
 power-law-like potential for the {\AE} models, it is still
 reasonable to consider such a potential phenomenologically.
 Second, the power-law-like potential given in Eq.~(\ref{eq9})
 subsumes the simple cases with {\em not only}
 $V(\theta, \phi)=f(\phi)\theta^2$ of~\cite{r30}
 and $V(\theta, \phi)=f(\theta^2)$ of~\cite{r23}, {\em but
 also} $V(\theta, \phi)=\frac{1}{2}m^2 \phi^2+\mu\theta\phi$
 of~\cite{r2}, in contrast to the exponential-like potential
 given in Eq.~(\ref{eq7}). Therefore, it is of interest to
 study the {\AE} models with this type of power-law-like potential.

Here, we focus on the cosmological evolution of the {\AE}
 models with the power-law-like potential proposed
 in Eq.~(\ref{eq9}). However, the potential is too complicated
 when $r$ is large. On the other hand, if we only consider the
 term $\theta f(\phi)$ (namely $r=1$), it will be canceled in
 $V-\theta V_\theta$, and then the energy density $\rho_{\phi}$
 and pressure $p_{\phi}$ in Eqs.~(\ref{eq3}) and (\ref{eq4})
 become trivial in this case. Therefore, in this work we
 consider the case with $r$ up to $2$, namely
 \be{eq10}
 V(\theta, \phi)=V_0\,\phi^n+\mu_1\theta\phi^{n/2}+\mu_2\theta^2\,.
 \ee
 Note that we consider a flat FRW universe
 and set $8\pi G=\hbar=c=1$ throughout this work. We use the
 method of dynamical system~\cite{r31} (see also
 e.g.~\cite{r32,r33,r34,r35}) to investigate the cosmological
 evolution of the {\AE} models with the power-law-like
 potential given in Eq.~(\ref{eq10}). There might be some
 scaling attractors in the dynamical system, and both the
 fractional densities of dark energy and matter
 are non-vanishing constants over there. The universe will
 eventually enter these scaling attractors regardless of the
 initial conditions, and hence it is not so surprising that we
 are living in an epoch in which the densities of dark energy
 and matter are comparable. Therefore, the cosmological coincidence
 problem could be alleviated without fine-tunings.


\section{The universe without matter}\label{sec2}

At first, we consider the cosmological evolution of the
 universe without matter, namely the total energy density is
 dominated by $\rho_\phi$. This case corresponds to the
 early universe dominated by inflaton, or the far future
 universe dominated by dark energy. The Lagrangians for
 the aether and scalar field read~\cite{r2}
 $${\cal L}_{\rm ae}=-\frac{M_p^2}{2}\left[R+
 K^{ab}_{\ \ \ cd}\nabla_a u^c \nabla_b u^d+
 \lambda\left(u^a u_a-1\right)\right]\,,~~~~~~~
 {\cal L}_\phi=\frac{1}{2}\nabla_a \phi \nabla^a \phi-
 V(\theta, \phi)\,,$$
 where $M_p$ is the reduced Planck mass, and
 $$K^{ab}_{\ \ \ cd}=
 c_1 g^{ab}g_{cd}+c_2\delta^a_c \delta^b_d+
 c_3\delta^a_d \delta^b_c+c_4 u^a u^b g_{cd}\,,$$
 in which $c_i$ are all dimensionless free parameters. After
 some algebra (see e.g.~\cite{r2,r5,r6}), one can find the
 stress tensor in Eq.~(\ref{eq2}), and then the corresponding
 energy density $\rho_{\phi}$ and pressure $p_{\phi}$ of the
 homogeneous scalar field in Eqs.~(\ref{eq3}) and (\ref{eq4}).
 Considering a flat FRW universe and using Eqs.~(\ref{eq3}),
 (\ref{eq4}), the corresponding Friedmann equation and
 Raychaudhuri equation are given
 by~\cite{r2,r5,r6}
 \bea
 &&\frac{1}{3}\theta^2=\rho_\phi=\frac{1}{2}\dot{\phi}^2+V
 -\theta V_\theta\,,\label{eq11}\\
 &&\frac{1}{3}\dot{\theta}=-\frac{1}{2}\left(
 \rho_\phi+p_\phi\right)=-\frac{1}{2}\left(\dot{\phi}^2+
 \dot{V}_\theta\right)\,.\label{eq12}
 \eea
 Following e.g.~\cite{r31,r32,r33,r34,r35}, we introduce
 three dimensionless variables
 \be{eq13}
 x_1\equiv\sqrt{\frac{3}{2\left(1+3\mu_2\right)}}\cdot
 \frac{\dot{\phi}}{\theta}\,,~~~~~~~
 x_2\equiv\sqrt{\frac{3V_0}{1+3\mu_1}}\cdot
 \frac{\phi^{n/2}}{\theta}\,,~~~~~~~
 x_3\equiv\phi^{-1}\,.
 \ee
 Then, the Friedmann equation~(\ref{eq11}) can be recast as
 \be{eq14}
 1=x_1^2+x_2^2\,.
 \ee
 From Eqs.~(\ref{eq11}), (\ref{eq12}) and (\ref{eq3}), (\ref{eq4}),
 we have
 \be{eq15}
 s\equiv-\frac{\dot{\theta}}{\theta^2}=x_1
 \left(x_1+\frac{1}{3}\nu_1 x_2 x_3\right)\,.
 \ee
 Note that for convenience we introduce two new constants
 \be{eq16}
 \nu_1\equiv\frac{3n\mu_1}{2\sqrt{2V_0}}\,,~~~~~~~
 \nu_2\equiv\sqrt{\frac{3\left(1+3\mu_2\right)}{2}}\,.
 \ee
 By the help of Eqs.~(\ref{eq11}), (\ref{eq12})
 and (\ref{eq3}), (\ref{eq4}), the evolution
 equation~(\ref{eq6}) can be recast as a dynamical system,
 namely
 \bea
 &&x_1^\prime=3\left(s-1\right)x_1-\left(
 \nu_1+n\nu_2 x_2\right)x_2 x_3\,,\label{eq17}\\
 &&x_2^\prime=\left(3s+n\nu_2 x_1 x_3 \right) x_2\,, \label{eq18}\\
 &&x_3^\prime=-2\nu_2 x_1 x_3^2\,,\label{eq19}
 \eea
 where $s$ is given in Eq.~(\ref{eq15}), and a prime denotes
 the derivative with respect to the so-called $e$-folding time
 $N\equiv\ln a$. On the other hand, using Eqs.~(\ref{eq11})
 and (\ref{eq12}), the equation-of-state parameter (EoS) is given by
 \be{eq20}
 w_\phi\equiv\frac{p_\phi}{\rho_\phi}=2s-1\,.
 \ee
 Of course, by definition, the deceleration parameter reads
 \be{eq21}
 q\equiv-\frac{\ddot{a}}{aH^2}=
 -1-3\frac{\dot{\theta}}{\theta^2}=3s-1\,.
 \ee
 We can obtain the critical points
 $(\bar{x}_1, \bar{x}_2, \bar{x}_3)$ of the above autonomous
 system by imposing the conditions
 $\bar{x}_1^\prime=\bar{x}_2^\prime=\bar{x}_3^\prime=0$ and
 the Friedmann constraint~(\ref{eq14}), i.e.,
 $\bar{x}_1^2+\bar{x}_2^2=1$. On the other hand,
 by definitions, $\bar{x}_1$, $\bar{x}_2$ and $\bar{x}_3$
 should be real. There are four critical points, and we
 present them in Table~\ref{tab1}. Points (D.1p)
 and (D.1m) correspond to a decelerated universe while the
 scalar field mimics a stiff fluid. Thus, they are not
 desirable in fact. Points (D.2p) and (D.2m) correspond to a
 de~Sitter universe while the scalar field mimics
 a cosmological constant. They are suitable to describe
 the inflationary universe in the early epoch, or
 the accelerated universe in the late time.

We study the stability of these critical points,
 by substituting the linear perturbations
 $x_1\to\bar{x}_1+\delta x_1$, $x_2\to\bar{x}_2+\delta x_2$,
 $x_3\to\bar{x}_3+\delta x_3$ about the critical point
 $(\bar{x}_1, \bar{x}_2, \bar{x}_3)$ into the autonomous
 system (\ref{eq17}), (\ref{eq18}), (\ref{eq19})
 and linearize them. Because of the Friedmann
 constraint~(\ref{eq14}), there are only two independent
 evolution equations, namely
 \bea
 &\disp\delta x_1^\prime=3\left[
 \left(\bar{s}-1\right)\delta x_1+\bar{x}_1\delta s\right]-
 \nu_1 \epsilon\sqrt{1-\bar{x}_1^2}\left(
 \delta x_3-\frac{\bar{x}_1\bar{x}_3\delta x_1}{1-\bar{x}_1^2}
 \right)-n\nu_2\left[\left(1-\bar{x}_1^2\right)
 \delta x_3-2\bar{x}_1 \bar{x}_3 \delta x_1\right]\,,~~~
 \label{eq22}\\[2mm]
 &\disp\delta x_3^\prime=-2\nu_2\left(\bar{x}_3^2 \delta x_1
 +2\bar{x}_1 \bar{x}_3 \delta x_3\right)\,,\label{eq23}
 \eea
 where $\epsilon$ is the sign of $\bar{x}_2$, and
 \bea
 &\disp\bar{s}=\bar{x}_1\left(\bar{x}_1+\frac{\bar{x}_3}{3}
 \nu_1\epsilon\sqrt{1-\bar{x}_1^2}\right)\,,\label{eq24}\\[2mm]
 &\disp\delta s=2\bar{x}_1\delta x_1+\frac{1}{3}\nu_1\epsilon
 \sqrt{1-\bar{x}_1^2}\left[\left(1-\frac{\bar{x}_1^2}{1-
 \bar{x}_1^2}\right)\bar{x}_3\delta x_1+\bar{x}_1\delta x_3
 \right]\,.\label{eq25}
 \eea
 The two eigenvalues of the coefficient matrix of
 Eqs.~(\ref{eq22}) and (\ref{eq23}) determine the stability of
 the critical point. The eigenvalues of both Points (D.1p) and
 (D.1m) are $\{6,~0\}$, and hence they are unstable. On the
 other hand, the eigenvalues of both Points (D.2p) and (D.2m)
 are $\{-3,~0\}$, and hence they are stable. So, in the case
 without matter, there are only two attractors (D.2p) and
 (D.2m), which correspond to an inflationary universe in the
 early epoch, or a de~Sitter universe in the late time.

\vspace{4mm}  


 \begin{table}[htbp]
 \begin{center}
 \begin{tabular}{l|c|c|c}
 \hline\hline & & & \\[-4mm]
 ~Label~ & ~~~Critical Point $(\bar{x}_1, \bar{x}_2, \bar{x}_3)$~~~
 & ~~~$w_\phi$~~~ & ~~~~$q$~~~~ \\[0.5mm]
 \hline & & & \\[-3mm]
 ~D.1p~ & ~ $1,\ 0,\ 0$ ~ & $1$ & $2$ \\[3mm]
 ~D.1m~ & ~ $-1,\ 0,\ 0$ ~ & $1$ & $2$ \\[3mm]
 ~D.2p~ & ~ $0,\ 1,\ 0$ ~ & $-1$ & $-1$ \\[3mm]
 ~D.2m~ & ~ $0,\ -1,\ 0$ ~ & $-1$ & $-1$\\[1.5mm]
 \hline\hline
 \end{tabular}
 \end{center}
 \caption{\label{tab1} Critical points for the autonomous
 system (\ref{eq17}), (\ref{eq18}), (\ref{eq19}) and their
 corresponding physical quantities.}
 \end{table}


\vspace{-2mm}  


\section{Adding matter}\label{sec3}

Here, we consider the universe containing also matter, which
 is described by a perfect fluid with barotropic EoS, namely
 \be{eq26}
 p_m=w_m\rho_m=\left(\gamma-1\right)\rho_m\,,
 \ee
 where the barotropic index $\gamma$ is a constant, and
 $0<\gamma<2$. In particular, $\gamma=1$ and $4/3$
 correspond to pressureless matter and radiation,
 respectively. In this case, the corresponding Friedmann
 equation and Raychaudhuri equation become
 \bea
 &&\frac{1}{3}\theta^2=\rho_\phi+\rho_m=\frac{1}{2}\dot{\phi}^2
 +V-\theta V_\theta+\rho_m\,,\label{eq27}\\
 &&\frac{1}{3}\dot{\theta}=-\frac{1}{2}\left(\rho_\phi+
 p_\phi+\rho_m+p_m\right)=-\frac{1}{2}\left(\dot{\phi}^2+
 \dot{V}_\theta+\gamma\rho_m\right)\,.\label{eq28}
 \eea
 The energy conservation equation of matter is given by
 \be{eq29}
 \dot{\rho}_m+\gamma\theta\rho_m=0\,.
 \ee
 In addition to $x_1$, $x_2$, $x_3$ defined in
 Eq.~(\ref{eq13}), we introduce another dimensionless variable
 \be{eq30}
 x_4\equiv\sqrt{\frac{3}{1+3\mu_2}}
 \cdot\frac{\sqrt{\rho_m}}{\theta}\,.
 \ee
 Now, the Friedmann equation~(\ref{eq27}) can be recast as
 \be{eq31}
 1=x_1^2+x_2^2+x_4^2\,.
 \ee
 From Eqs.~(\ref{eq27}), (\ref{eq28})
 and (\ref{eq3}), (\ref{eq4}), we find that
 \be{eq32}
 s\equiv-\frac{\dot{\theta}}{\theta^2}=x_1^2
 +\frac{1}{3}\nu_1 x_1 x_2 x_3+\frac{\gamma}{2}x_4^2\,.
 \ee
 By the help of Eqs.~(\ref{eq27}), (\ref{eq28})
 and (\ref{eq3}), (\ref{eq4}), the evolution
 equations~(\ref{eq6}) and (\ref{eq29}) can be recast as
 a dynamical system, namely
 \bea
 &&x_1^\prime=3\left(s-1\right)x_1-\left(
 \nu_1+n\nu_2 x_2\right)x_2 x_3\,,\label{eq33}\\
 &&x_2^\prime=\left(3s+n\nu_2 x_1 x_3 \right) x_2\,,\label{eq34}\\
 &&x_3^\prime=-2\nu_2 x_1 x_3^2\,,\label{eq35}\\
 &&x_4^\prime=3 x_4\left(s-\frac{\gamma}{2} \right)\,,\label{eq36}
 \eea
 where $s$ is given in Eq.~(\ref{eq32}). On the other hand,
 using Eqs.~(\ref{eq3}), (\ref{eq31}) and (\ref{eq16}),
 it is easy to find that the fractional energy densities
 $\Omega_i\equiv 3\rho_i/\theta^2$ of the scalar field and
 matter are given by
 \be{eq37}
 \Omega_\phi=\left(1+3\mu_2\right)\left(x_1^2+x_2^2\right)-
 3\mu_2=1-\frac{2}{3}\nu_2^2 x_4^2\,,~~~~~~~
 \Omega_m=\frac{2}{3}\nu_2^2 x_4^2\,,
 \ee
 and they satisfy $\Omega_\phi+\Omega_m=1$. Note that due to
 the $\mu_2 \theta^2$ term in the potential $V(\theta, \phi)$
 (see Eq.~(\ref{eq10})), one {\em cannot} naively write
 $\Omega_\phi=x_1^2+x_2^2$ and $\Omega_m=x_4^2$. Using
 Eqs.~(\ref{eq27}) and (\ref{eq28}), the total EoS reads
 \be{eq38}
 w_{\rm tot}\equiv\frac{p_{\rm tot}}{\rho_{\rm tot}}=2s-1\,.
 \ee
 Noting that $w_{\rm tot}\equiv p_{\rm tot}/\rho_{\rm tot}=
 \Omega_\phi w_\phi+\Omega_m w_m$, we find the EoS of the
 scalar field as
 \be{eq39}
 w_\phi\equiv\frac{p_\phi}{\rho_\phi}=\frac{w_{\rm tot}
 -\Omega_m\left(\gamma-1\right)}{\Omega_\phi}=
 \gamma-1+\frac{3\left(2s-\gamma\right)}{3-2\nu_2^2 x_4^2}\,.
 \ee
 By definition, the deceleration parameter $q$ is the same
 in Eq.~(\ref{eq21}), but in which $s$ has been changed to
 the one in Eq.~(\ref{eq32}). We can obtain the critical
 points $(\bar{x}_1, \bar{x}_2, \bar{x}_3, \bar{x}_4)$ of
 the above autonomous system by imposing the conditions
 $\bar{x}_1^\prime=\bar{x}_2^\prime=
 \bar{x}_3^\prime=\bar{x}_4^\prime=0$ and the
 Friedmann constraint~(\ref{eq31}), i.e.,
 $\bar{x}_1^2+\bar{x}_2^2+x_4^2=1$.
 On the other hand, by definitions, $\bar{x}_1$, $\bar{x}_2$
 $\bar{x}_3$, $\bar{x}_4$ should be real, and
 $\bar{x}_4\geq 0$. There are five critical points, and we
 present them in Table~\ref{tab2}. Points (M.1p) and (M.1m)
 correspond to a decelerated universe while the
 scalar field mimics a stiff fluid. Points (M.2p) and (M.2m)
 correspond to a de~Sitter universe while the scalar field
 mimics a cosmological constant. Note that these
 four solutions all correspond to a universe dominated by
 the scalar field (dark energy). Point (M.3) is a scaling
 solution if $\nu_2^2\not=0$ or $3/2$. Since its
 $q=3\gamma/2-1$, the universe is decelerated ($q>0$) if
 $\gamma>2/3$, and is accelerated ($q<0$) if $\gamma<2/3$.
 So, in the case of $\gamma=1$ (pressureless matter) and
 $\gamma=4/3$ (radiation), the universe cannot be accelerated.

To study the stability of these critical points, we substitute
 the linear perturbations $x_1\to\bar{x}_1+\delta x_1$,
 $x_2\to\bar{x}_2+\delta x_2$, $x_3\to\bar{x}_3+\delta x_3$,
 $x_4\to\bar{x}_4+\delta x_4$ about the critical point
 $(\bar{x}_1, \bar{x}_2, \bar{x}_3, \bar{x}_4)$ into the
 autonomous system (\ref{eq33})---(\ref{eq36}) and linearize
 them. Because of the Friedmann constraint~(\ref{eq31}),
 there are only three independent evolution equations, namely
 \bea
 \disp\delta x_1^\prime &=& 3\left[
 \left(\bar{s}-1\right)\delta x_1+\bar{x}_1\delta s\right]-
 \nu_1 \epsilon\sqrt{1-\bar{x}_1^2-\bar{x}_4^2}\left[\delta x_3
 -\frac{\bar{x}_3\left(\bar{x}_1\delta x_1+\bar{x}_4\delta x_4
 \right)}{1-\bar{x}_1^2-\bar{x}_4^2}\right]\nonumber\\[1mm]
 & &-n\nu_2\left[\left(1-\bar{x}_1^2-\bar{x}_4^2\right)
 \delta x_3-2\bar{x}_3\left(\bar{x}_1 \delta x_1+
 \bar{x}_4\delta x_4\right)\right]\,,\label{eq40}\\[2mm]
 \disp\delta x_3^\prime &=& -2\nu_2\left(\bar{x}_3^2 \delta x_1
 +2\bar{x}_1 \bar{x}_3 \delta x_3\right)\,,\label{eq41}\\[2mm]
 \disp\delta x_4^\prime &=& 3\left(
 \bar{x}_4\delta s+\bar{s}\delta x_4\right)\,,\label{eq42}
 \eea
 where $\epsilon$ is the sign of $\bar{x}_2$, and
 \bea
 &\disp\bar{s}=\bar{x}_1^2+\frac{1}{3}\nu_1\bar{x}_1\bar{x}_3\,
 \epsilon\sqrt{1-\bar{x}_1^2-\bar{x}_4^2}+
 \frac{\gamma}{2}\bar{x}_4^2\,,\label{eq43}\\[2mm]
 &\disp\delta s=2\bar{x}_1\delta x_1+\gamma\bar{x}_4\delta x_4
 +\frac{1}{3}\nu_1\epsilon\sqrt{1-\bar{x}_1^2-\bar{x}_4^2}
 \left[\bar{x}_1\delta x_3+\bar{x}_3\delta x_1-
 \frac{\bar{x}_1\bar{x}_3\left(\bar{x}_1\delta x_1
 +\bar{x}_4\delta x_4\right)}{1-\bar{x}_1^2-\bar{x}_4^2}
 \right]\,.\label{eq44}
 \eea
 The three eigenvalues of the coefficient matrix of
 Eqs.~(\ref{eq40})---(\ref{eq42}) determine the stability of
 the critical point. The eigenvalues of both Points (M.1p) and
 (M.1m) are $\{6,~3,~0\}$, and hence they are unstable. On the
 other hand, the eigenvalues of both Points (M.2p) and (M.2m)
 are $\{-3,~0,~0\}$, and hence they are stable. The eigenvalues
 of Point (M.3) is $\{9\gamma/2,~3(\gamma-2)/2,~0\}$, and
 hence it is unstable, since $\gamma$ is positive. So, for the
 case with matter, there are only two attractors (M.2p) and
 (M.2m), which correspond to an inflationary universe in the
 early epoch, or a de~Sitter universe in the late time.
 Unfortunately, there is {\em no} scaling attractor in this
 case, since Point (M.3) is unstable. However,
 the scaling attractor is necessary to alleviate
 the cosmological coincidence problem. So, we should try to
 find a way out.

\vspace{4mm}  


 \begin{table}[htbp]
 \begin{center}
 \begin{tabular}{l|c|c|c|c|c|c}
 \hline\hline & & & & & &\\[-4mm]
 ~Label~ & ~~~Critical Point
 $(\bar{x}_1, \bar{x}_2, \bar{x}_3, \bar{x}_4)$~~~
 & ~~~$\Omega_\phi$~~~ & ~~~$\Omega_m$~~~ & ~~~$w_{\rm tot}$~~~
 & ~~~$w_\phi$~~~ & ~~~~$q$~~~~ \\[0.5mm]
 \hline & & & & & & \\[-3mm]
 ~M.1p~ & ~ $1,\ 0,\ 0,\ 0$ ~ & $1$ & $0$ & $1$ & $1$ & $2$ \\[3mm]
 ~M.1m~ & ~ $-1,\ 0,\ 0,\ 0$ ~ & $1$ & $0$ & $1$ & $1$ & $2$\\[3mm]
 ~M.2p~ & ~ $0,\ 1,\ 0,\ 0$ ~ & $1$ & $0$ & $-1$ & $-1$ & $-1$\\[3mm]
 ~M.2m~ & ~ $0,\ -1,\ 0,\ 0$ ~ & $1$ & $0$ & $-1$ & $-1$ & $-1$\\[3mm]
 ~M.3~ & ~ $0,\ 0,\ {\rm any},\ 1$ ~
 & ~~$\disp 1-\frac{2}{3}\nu_2^2$~~ & $\disp\frac{2}{3}\nu_2^2$
 & $\gamma-1$ & $\gamma-1$ & ~~$\disp \frac{3}{2}\gamma-1$~ \\[3mm]
 \hline\hline
 \end{tabular}
 \end{center}
 \caption{\label{tab2} Critical points for the autonomous
 system (\ref{eq33})---(\ref{eq36}) and their corresponding
 physical quantities.}
 \end{table}


\vspace{-1.5mm}  


\section{Allowing the interaction between dark energy and
 matter}\label{sec4}

As mentioned in the previous section, we should try to find
 a suitable way to alleviate the cosmological coincidence
 problem. In the literature, the common way is allowing the
 interaction between dark energy and matter. If dark energy
 can decay into matter (vice versa), these two components
 might achieve a balance and then their fractional energy
 densities become constant at some scaling attractors (if any).
 This is the key to alleviate the cosmological coincidence
 problem. Therefore, we allow the interaction between dark
 energy and matter in this section.

We assume that dark energy and matter interact through a
 coupling term $Q$, according to
 \bea
 &&\dot{\rho}_\phi+\theta\left(\rho_\phi+p_\phi\right)=
 -Q\,,\label{eq45}\\
 &&\dot{\rho}_m+\theta\left(\rho_m+p_m\right)=Q\,,\label{eq46}
 \eea
 which preserves the total energy conservation equation
 $\dot{\rho}_{\rm tot}+\theta\left(\rho_{\rm tot}
 +p_{\rm tot}\right)=0$. Due to the interaction $Q$ in
 Eq.~(\ref{eq45}), the equation of motion should be modified
 accordingly, i.e.,
 \be{eq47}
 \ddot{\phi}+\theta\dot{\phi}+V_\phi=-\frac{Q}{\dot{\phi}}\,.
 \ee
 By the help of Eqs.~(\ref{eq27}), (\ref{eq28})
 and (\ref{eq3}), (\ref{eq4}), the evolution
 equations~(\ref{eq45}) and (\ref{eq46}) can be recast as
 a dynamical system, namely
 \bea
 &&x_1^\prime=3\left(s-1\right)x_1-
 \left(\nu_1+n\nu_2 x_2\right)x_2 x_3-Q_1\,,\label{eq48}\\
 &&x_2^\prime=\left(3s+n\nu_2 x_1 x_3 \right) x_2\,,\label{eq49}\\
 &&x_3^\prime=-2\nu_2 x_1 x_3^2\,,\label{eq50}\\
 &&x_4^\prime=3x_4\left(s-\frac{\gamma}{2}\right)+Q_2\,,\label{eq51}
 \eea
 where $s$ is given in Eq.~(\ref{eq32}), and
 \be{eq52}
 Q_1\equiv\sqrt{\frac{3}{2\left(1+3\mu_2\right)}}
 \cdot\frac{3Q}{\theta^2 \dot{\phi}}\,,~~~~~~~
 Q_2\equiv\frac{3x_4 Q}{2\theta\rho_m}\,.
 \ee
 Eqs.~(\ref{eq48})---(\ref{eq51}) could be an autonomous system
 if the interaction term $Q$ is chosen to be suitable forms.
 In the present work, we will consider four most familiar
 interaction terms extensively considered in the literature
 (see e.g.~\cite{r31,r32,r33,r34,r35}), namely,
 Case (I) $Q=\eta_1\rho_m\dot{\phi}$, Case (II)
 $Q=\eta_2\theta\rho_m$, Case (III) $Q=\eta_3\theta\rho_\phi$,
 and Case (IV) $Q=\eta_4\theta\rho_{\rm tot}$, where $\eta_i$
 are all constants. Once the interaction term $Q$ is specified,
 we can obtain the critical points
 $(\bar{x}_1, \bar{x}_2, \bar{x}_3, \bar{x}_4)$ of the above
 autonomous system by imposing the conditions $\bar{x}_1^\prime
 =\bar{x}_2^\prime=\bar{x}_3^\prime=\bar{x}_4^\prime=0$ and
 the Friedmann constraint~(\ref{eq31}), i.e.,
 $\bar{x}_1^2+\bar{x}_2^2+x_4^2=1$. On the other hand, by
 definitions, $\bar{x}_1$, $\bar{x}_2$ $\bar{x}_3$, $\bar{x}_4$
 should be real, and $\bar{x}_4\geq 0$. The physical quantities
 of the critical points, namely $\Omega_\phi$, $\Omega_m$,
 $w_{\rm tot}$, $w_\phi$ and $q$, are given in
 Eqs.~(\ref{eq37}), (\ref{eq38}), (\ref{eq39}), (\ref{eq21}),
 and in which $s$ is given in Eq.~(\ref{eq32}).

Once the critical points are available, their stability should
 be investigated. To this end, we substitute the linear
 perturbations $x_1\to\bar{x}_1+\delta x_1$,
 $x_2\to\bar{x}_2+\delta x_2$, $x_3\to\bar{x}_3+\delta x_3$,
 $x_4\to\bar{x}_4+\delta x_4$ about the critical
 point $(\bar{x}_1, \bar{x}_2, \bar{x}_3, \bar{x}_4)$ into the
 autonomous system (\ref{eq48})---(\ref{eq51}) and linearize
 them. Because of the Friedmann constraint~(\ref{eq31}),
 there are only three independent evolution equations, namely
 \bea
 \disp\delta x_1^\prime &=& 3\left[
 \left(\bar{s}-1\right)\delta x_1+\bar{x}_1\delta s\right]-
 \nu_1 \epsilon\sqrt{1-\bar{x}_1^2-\bar{x}_4^2}\left[\delta x_3
 -\frac{\bar{x}_3\left(\bar{x}_1\delta x_1+\bar{x}_4\delta x_4
 \right)}{1-\bar{x}_1^2-\bar{x}_4^2}\right]\nonumber\\[1mm]
 & &-n\nu_2\left[\left(1-\bar{x}_1^2-\bar{x}_4^2\right)
 \delta x_3-2\bar{x}_3\left(\bar{x}_1 \delta x_1+
 \bar{x}_4\delta x_4\right)\right]-\delta Q_1\,,\label{eq53}\\[2mm]
 \disp\delta x_3^\prime &=& -2\nu_2\left(\bar{x}_3^2 \delta x_1
 +2\bar{x}_1 \bar{x}_3 \delta x_3\right)\,,\label{eq54}\\[2mm]
 \disp\delta x_4^\prime &=& 3\left(\bar{x}_4\delta s+\bar{s}
 \delta x_4\right)+\delta Q_2\,,\label{eq55}
 \eea
 where $\epsilon$ is the sign of $\bar{x}_2$, and $\delta Q_1$,
 $\delta Q_2$ are the linear perturbations coming from $Q_1$,
 $Q_2$, respectively. Note that $\bar{s}$ and $\delta s$ are
 given in Eqs.~(\ref{eq43}) and (\ref{eq44}). The three
 eigenvalues of the coefficient matrix of
 Eqs.~(\ref{eq53})---(\ref{eq55}) determine the stability of
 the critical point.


\subsection{Case (I) $Q=\eta_1\rho_m\dot{\phi}$}\label{sec4a}

At first, we consider the case with $Q=\eta_1\rho_m\dot{\phi}$.
 The corresponding $Q_1=\eta_1\nu_2 x_4^2$ and
 $Q_2=\eta_1\nu_2 x_1 x_4$. In this case, there are five
 critical points, and we present them in Table~\ref{tab3}.
 Points (I.1p) and (I.1m) correspond to a decelerated universe
 while the scalar field mimics a stiff fluid. Points (I.2p)
 and (I.2m) correspond to a de~Sitter universe while the
 scalar field mimics a cosmological constant. Note that these
 four solutions all correspond to a universe dominated by
 the scalar field (dark energy). Point (I.3) is a scaling
 solution, and its physical quantities read
 \bea
 &\disp\Omega_\phi=1-\frac{2}{3}\nu_2^2+\frac{8\eta_1^2
 \nu_2^4}{27(\gamma-2)^2}\,,~~~~~~~
 \Omega_m=\frac{2}{3}\nu_2^2-\frac{8\eta_1^2
 \nu_2^4}{27(\gamma-2)^2}\,,\nonumber\\[2mm]
 &\disp w_{\rm tot}=\gamma-1-\frac{4\eta_1^2
 \nu_2^2}{9(\gamma-2)}\,,~~~~~~~
 w_\phi=\gamma-1+\frac{12(\gamma-2)\eta_1^2\nu_2^2}{(18
 \nu_2^2-27)(\gamma-2)^2-8\eta_1^2\nu_2^4}\,,\nonumber\\[2mm]
 &\disp q=\frac{3}{2}\gamma-
 1+\frac{2\eta_1^2\nu_2^2}{3(2-\gamma)}\,.\label{eq56}
 \eea
 Since $0<\gamma<2$, it is worth noting that at Point (I.3)
 the universe is decelerated ($q>0$) when $\gamma>2/3$.
 Unfortunately, the universe cannot be accelerated
 if $\gamma=1$ (pressureless matter) and $4/3$ (radiation).

We turn to the stability of these critical points. The corresponding
 $\delta Q_1=2\eta_1\nu_2\bar{x}_4\delta x_4$ and
 $\delta Q_2=\eta_1\nu_2(\bar{x}_1\delta x_4+\bar{x}_4\delta x_1)$.
 The eigenvalues of Point (I.1p) are $\{6,~0,~3+\eta_1\nu_2\}$,
 and hence it is unstable. The eigenvalues of Point (I.1m) are
 $\{6,~0,~3-\eta_1\nu_2\}$, and hence it is also unstable. On
 the other hand, the eigenvalues of both Points (I.2p) and
 (I.2m) are $\{-3,~0,~0\}$. So, they are stable, and hence
 they are de~Sitter attractors in fact. The eigenvalues of
 Point (I.3) are
 \be{eq57}
 \left\{0,~-\frac{3}{2}+3\gamma+\frac{\eta_1^2\nu_2^2}{2-
 \gamma}-\sigma_1,~-\frac{3}{2}+3\gamma+\frac{\eta_1^2\nu_2^2}
 {2-\gamma}+\sigma_1\right\}\,,
 \ee
 where
 \be{eq58}
 \sigma_1\equiv\frac{1}{2}\left[9(1+\gamma)^2-\frac{4\left(\gamma^2
 +3\gamma-2\right)\eta_1^2\nu_2^2}{(\gamma-2)^2}+
 \frac{4\eta_1^4\nu_2^4}{9(\gamma-2)^2}\right]^{1/2}\,.
 \ee
 Point (I.3) can exist and is stable under some conditions,
 which are too complicated to be presented here. Nevertheless,
 we would like to say more. Since $0<\gamma<2$, it is worth
 noting that if $\gamma>1/2$, at least one of its eigenvalues
 is positive (when $\sigma_1$ is a real number), or the real
 parts of both last two eigenvalues are positive
 (when $\sigma_1$ is an imaginary number). Therefore, Point
 (I.3) is certainly unstable for $\gamma>1/2$.
 So, if $\gamma=1$ (pressureless matter) and $4/3$ (radiation),
 Point (I.3) cannot be an attractor. It can be a scaling
 attractor only when $\gamma<1/2$ (necessary but not sufficient
 condition), in this case the EoS of
 matter $w_m=\gamma-1<-1/2$, which is dark energy in fact.

\vspace{4mm}  


 \begin{table}[htbp]
 \begin{center}
 \begin{tabular}{l|c|ccccc}
 \hline\hline & & & & & &\\[-4mm]
 ~Label~ & ~~~Critical Point
 $(\bar{x}_1, \bar{x}_2, \bar{x}_3, \bar{x}_4)$~~~
 & ~~~$\Omega_\phi$~~~ & ~~~$\Omega_m$~~~ & ~~~$w_{\rm tot}$~~~
 & ~~~$w_\phi$~~~ & ~~~~$q$~~~~ \\[0.5mm]
 \hline & & & & & & \\[-3mm]
 ~I.1p~ & ~ $1,\ 0,\ 0,\ 0$ ~ & $1$ & $0$ & $1$ & $1$ & $2$ \\[3mm]
 ~I.1m~ & ~ $-1,\ 0,\ 0,\ 0$ ~ & $1$ & $0$ & $1$ & $1$ & $2$\\[3mm]
 ~I.2p~ & ~ $0,\ 1,\ 0,\ 0$ ~ & $1$ & $0$ & $-1$ & $-1$ & $-1$\\[3mm]
 ~I.2m~ & ~ $0,\ -1,\ 0,\ 0$ ~ & $1$ & $0$ & $-1$ & $-1$ & $-1$\\[3mm]
 ~I.3~ & ~ $\disp\frac{2\eta_1\nu_2}{3(\gamma-2)},\ 0,\ 0,
  \ \sqrt{1-\frac{4\eta_1^2\nu_2^2}{9(\gamma-2)^2}}$ ~
 & & & Eq.~(\ref{eq56}) & & \\[4mm]
 \hline\hline
 \end{tabular}
 \end{center}
 \caption{\label{tab3} Critical points for the autonomous
 system (\ref{eq48})---(\ref{eq51}) and their corresponding
 physical quantities,
 in the Case (I) $Q=\eta_1\rho_m\dot{\phi}$.}
 \end{table}


\vspace{-6.4mm}  


\subsection{Case (II) $Q=\eta_2\theta\rho_m$}\label{sec4b}

The second choice of interaction is $Q=\eta_2\theta\rho_m$.
 The corresponding $Q_1=\frac{3}{2}\eta_2 x_4^2 x_1^{-1}$
 and $Q_2=\frac{3}{2}\eta_2 x_4$. In this case, there are four
 critical points, and we present them in Table~\ref{tab4}.
 Points (II.1p) and (II.1m) correspond to a decelerated
 universe while the scalar field mimics a stiff fluid. Note
 that these two solutions both correspond to a universe
 dominated by the scalar field (dark energy). Points (II.2p)
 and (II.2m) are scaling solutions, and their physical
 quantities read
 \bea
 &\disp \Omega_\phi=1-\frac{2}{3}\nu_2^2\left(1-
 \frac{\eta_2}{\gamma-2}\right)\,,~~~~~~~
 \Omega_m=\frac{2}{3}\nu_2^2\left(1-\frac{\eta_2}{\gamma
 -2}\right)\,,\nonumber\\[2mm]
 &\disp w_{\rm tot}=\gamma-1-\eta_2\,,~~~~~~~
 w_\phi=\gamma-1+\frac{3(\gamma-2)\eta_2}{6-
 3\gamma+2(\gamma-2-\eta_2)\nu_2^2}\,,\nonumber\\[1mm]
 &\disp q=\frac{1}{2}\left(3\gamma-2-3\eta_2
 \right)\,.\label{eq59}
 \eea
 Since $0<\gamma<2$, both Points (II.2p) and (II.2m) can
 exist under the conditions $\gamma-2\leq\eta_2\leq 0$ and
 $0\leq\Omega_m\leq 1$. From Eq.~(\ref{eq59}), the condition
 to accelerate the universe ($q<0$) is $\gamma-\eta_2<2/3$.
 On the other hand, if $\gamma>2/3$, the universe
 is certainly decelerated ($q>0$) because $\eta_2\leq 0$
 is required by the existence of Points (II.2p) and (II.2m).
 Unfortunately, the universe cannot be accelerated
 if $\gamma=1$ (pressureless matter) and $4/3$ (radiation).

We then consider the stability of these critical points. The
 corresponding $\delta Q_1=\frac{3}{2}\eta_2\bar{x}_4\bar{x}_1^{-1}
 (2\delta x_4-\bar{x}_4\bar{x}_1^{-1}\delta x_1)$ and
 $\delta Q_2=\frac{3}{2}\eta_2\delta x_4$. The eigenvalues of
 both Points (II.1p) and (II.1m) are $\{6,~0,~3(\eta_2+2)/2\}$,
 and hence they are unstable. The eigenvalues of
 both Points (II.2p) and (II.2m) are
 \be{eq60}
 \left\{0,~-3+\frac{15}{4}\gamma-3\eta_2-\sigma_2,~-3+
 \frac{15}{4}\gamma-3\eta_2+\sigma_2\right\}\,,
 \ee
 where
 \be{eq61}
 \sigma_2\equiv\frac{3}{4}\left[(\gamma+4)^2
 -\frac{16\gamma\eta_2}{\gamma-2}\right]^{1/2}\,.
 \ee
 Points (II.2p) and (II.2m) can exist and are stable under some
 conditions, which are too complicated to be presented here.
 Nevertheless, we would like to say more. Since
 $\gamma-2\leq\eta_2\leq 0$ is required by the existence of
 Points (II.2p) and (II.2m), $\sigma_2$ is a positive real
 number. On the other hand, if $\gamma>4/5$, at least one of
 the last two eigenvalues is positive. Therefore, Points
 (II.2p) and (II.2m) is certainly unstable for $\gamma>4/5$.
 So, if $\gamma=1$ (pressureless matter) and $4/3$ (radiation),
 Points (II.2p) and (II.2m) cannot be attractors. They can
 be scaling attractors only when $\gamma<4/5$ (necessary
 but not sufficient condition). As mentioned
 above, $\gamma<2/3+\eta_2<2/3$ is required to accelerate the
 universe at Points (II.2p) and (II.2m). In this case, the
 EoS of matter $w_m=\gamma-1<-1/3$, which is dark energy in fact.

\vspace{5.5mm}  


 \begin{table}[htbp]
 \begin{center}
 \begin{tabular}{l|c|ccccc}
 \hline\hline & & & & & &\\[-4mm]
 ~Label~ & ~~~Critical Point
 $(\bar{x}_1, \bar{x}_2, \bar{x}_3, \bar{x}_4)$~~~
 & ~~~$\Omega_\phi$~~~ & ~~~$\Omega_m$~~~ & ~~~$w_{\rm tot}$~~~
 & ~~~$w_\phi$~~~ & ~~~~$q$~~~~ \\[0.5mm]
 \hline & & & & & & \\[-3mm]
 ~II.1p~ & ~ $1,\ 0,\ 0,\ 0$ ~ & $1$ & $0$ & $1$ & $1$ & $2$ \\[3mm]
 ~II.1m~ & ~ $-1,\ 0,\ 0,\ 0$ ~ & $1$ & $0$ & $1$ & $1$ & $2$\\[3mm]
 ~II.2p~ & ~ $\disp\sqrt{\frac{\eta_2}{\gamma-2}},\ 0,\ 0,
 \ \sqrt{1-\frac{\eta_2}{\gamma-2}}$ ~ & &
 & Eq.~(\ref{eq59}) & &\\[4mm]
 ~II.2m~ & ~ $\disp-\sqrt{\frac{\eta_2}{\gamma-2}},\ 0,\ 0,
 \ \sqrt{1-\frac{\eta_2}{\gamma-2}}$ ~ & &
 & Eq.~(\ref{eq59}) & &\\[4mm]
 \hline\hline
 \end{tabular}
 \end{center}
 \caption{\label{tab4} Critical points for the autonomous
 system (\ref{eq48})---(\ref{eq51}) and their corresponding
 physical quantities, in the
 Case (II) $Q=\eta_2\theta\rho_m$.}
 \end{table}


\vspace{-3mm}  


 \begin{table}[htbp]
 \begin{center}
 \begin{tabular}{l|c}
 \hline\hline &\\[-4mm]
 ~Label~ & ~~~Critical Point
 $(\bar{x}_1, \bar{x}_2, \bar{x}_3, \bar{x}_4)$~~~\\[0.5mm]
 \hline & \\[-3mm]
 ~III.1p~ & ~ $\disp\sqrt{\frac{1}{2}\left(1-\frac{\eta_3}{\gamma-2}
 +\sigma_3\right)},\ 0,\ 0,\ \sqrt{\frac{1}{2}\left(1+\frac{\eta_3}
 {\gamma-2}-\sigma_3\right)}$ ~ \\[5mm]
 ~III.1m~ & ~ $\disp-\sqrt{\frac{1}{2}\left(1-\frac{\eta_3}{\gamma-2}
 +\sigma_3\right)},\ 0,\ 0,\ \sqrt{\frac{1}{2}\left(1+\frac{\eta_3}
 {\gamma-2}-\sigma_3\right)}$ ~ \\[5mm]
 ~III.2p~ & ~ $\disp\sqrt{\frac{1}{2}\left(1-\frac{\eta_3}{\gamma-2}
 -\sigma_3\right)},\ 0,\ 0,\ \sqrt{\frac{1}{2}\left(1+\frac{\eta_3}
 {\gamma-2}+\sigma_3\right)}$ ~ \\[5mm]
 ~III.2m~ & ~ $\disp-\sqrt{\frac{1}{2}\left(1-\frac{\eta_3}{\gamma-2}
 -\sigma_3\right)},\ 0,\ 0,\ \sqrt{\frac{1}{2}\left(1+\frac{\eta_3}
 {\gamma-2}+\sigma_3\right)}$ ~ \\[4mm]
 \hline\hline
 \end{tabular}
 \end{center}
 \caption{\label{tab5} Critical points for the autonomous
 system (\ref{eq48})---(\ref{eq51}), in the Case
 (III) $Q=\eta_3\theta\rho_\phi$.}
 \end{table}



\subsection{Case (III) $Q=\eta_3\theta\rho_\phi$}\label{sec4c}

The third choice of interaction is $Q=\eta_3\theta\rho_\phi$.
 The corresponding $Q_1=\frac{9}{4}\eta_3\nu_2^{-2}\left(1-
 \frac{2}{3}\nu_2^2 x_4^2\right)x_1^{-1}$ and $Q_2=\frac{9}{4}
 \eta_3\nu_2^{-2}x_4^{-1}-\frac{3}{2}\eta_3 x_4$.
 In this case, there are eight critical points. The first four
 critical points can exist only when $\eta_3=0$, i.e., there is
 no interaction between dark energy and matter. So, they reduce
 to the cases considered in Sec.~\ref{sec3}, and here we do not
 consider them any more. We present the other four critical
 points in Table~\ref{tab5}. Note that we have introduced
 a new constant
 \be{eq62}
 \sigma_3\equiv\frac{\sqrt{(\gamma-2+\eta_3)^2 \nu_2^2
 -6(\gamma-2)\eta_3}}{(\gamma-2)\nu_2}\,.
 \ee
 Points (III.1p) and (III.1m) are scaling solutions, and
 their physical quantities read
 \bea
 &\disp\Omega_\phi=1-\frac{\nu_2^2}{3}\left(1-\sigma_3+
 \frac{\eta_3}{\gamma-2}\right)\,,~~~~~~~
 \Omega_m=\frac{\nu_2^2}{3}\left(1-\sigma_3+
 \frac{\eta_3}{\gamma-2}\right)\,,\nonumber\\[2mm]
 &\disp w_{\rm tot}=\frac{1}{2}\left[\gamma+
 \eta_3+(2-\gamma)\sigma_3\right]\,,~~~~~~~
 w_\phi=\gamma-1-\frac{3(\gamma-2)\left[(\gamma-2)(1+\sigma_3)
 -\eta_3\right]}{2(\gamma-2)\left(3-\nu_2^2+
 \nu_2^2\sigma_3\right)-2\eta_3\nu_2^2}\,,\nonumber\\[1mm]
 &\disp q=\frac{1}{4}\left[2+3\gamma+3\eta_3
 -3\sigma_3(\gamma-2)\right]\,.\label{eq63}
 \eea
 Since $0<\gamma<2$, they can exist under the conditions
 $2-\gamma\geq\eta_3-\sigma_3(\gamma-2)\geq\gamma-2$ and
 $0\leq\Omega_m\leq 1$, as well as $\sigma_3$ is real.
 Therefore,
 $q\geq\left[2+3\gamma+3(\gamma-2)\right]/4=(3\gamma-2)/2$.
 So, if $\gamma>2/3$, the universe is certainly decelerated
 ($q>0$). Unfortunately, the universe cannot be accelerated
 if $\gamma=1$ (pressureless matter) and $4/3$ (radiation).
 Points (III.2p) and (III.2m) are also scaling solutions, and
 their physical quantities read
 \bea
 &\disp\Omega_\phi=1-\frac{\nu_2^2}{3}\left(1+\sigma_3+
 \frac{\eta_3}{\gamma-2}\right)\,,~~~~~~~
 \Omega_m=\frac{\nu_2^2}{3}\left(1+\sigma_3+
 \frac{\eta_3}{\gamma-2}\right)\,,\nonumber\\[2mm]
 &\disp w_{\rm tot}=\frac{1}{2}\left[\gamma+
 \eta_3+(\gamma-2)\sigma_3\right]\,,~~~~~~~
 w_\phi=\gamma-1-\frac{3(\gamma-2)\left[(\gamma-2)(1-\sigma_3)
 -\eta_3\right]}{2(\gamma-2)\left(3-\nu_2^2-
 \nu_2^2\sigma_3\right)-2\eta_3\nu_2^2}\,,\nonumber\\[1mm]
 &\disp q=\frac{1}{4}\left[2+3\gamma+3\eta_3
 +3\sigma_3(\gamma-2)\right]\,.\label{eq64}
 \eea
 Since $0<\gamma<2$, they can exist under the conditions
 $2-\gamma\geq\eta_3+\sigma_3(\gamma-2)\geq\gamma-2$ and
 $0\leq\Omega_m\leq 1$, as well as $\sigma_3$ is real.
 Therefore,
 $q\geq\left[2+3\gamma+3(\gamma-2)\right]/4=(3\gamma-2)/2$.
 So, if $\gamma>2/3$, the universe is certainly decelerated
 ($q>0$). Unfortunately, the universe cannot be accelerated
 if $\gamma=1$ (pressureless matter) and $4/3$ (radiation).

Then, we consider the stability of these critical points. In
 this case, we find the corresponding $\delta Q_1=-\frac{9}{4}\eta_3
 \nu_2^{-2}\left[\left(1-\frac{2}{3}\nu_2^2\bar{x}_4^2\right)
 \bar{x}_1^{-2}\delta x_1+\frac{4}{3}\nu_2^2\bar{x}_4\bar{x}_1^{-1}
 \delta x_4\right]$ and $\delta Q_2=-\frac{3}{2}\eta_3
 \delta x_4\left(1+\frac{3}{2}\nu_2^{-2}\bar{x}_4^{-2}\right)$.
 The eigenvalues of both Points (III.1p) and (III.1m) are
 \be{eq65}
 \left\{0,~\frac{3}{4}\left[\,2+2\gamma+\eta_3
 -3(\gamma-2)\sigma_3-\sqrt{\sigma_{31}}\, \right]\,,~
 \frac{3}{4}\left[\,2+2\gamma+\eta_3-3(\gamma-2)\sigma_3
 +\sqrt{\sigma_{31}} \, \right]\right\}\,,
 \ee
 where $\sigma_{31}$ is a function of $\gamma$, $\eta_3$ and
 $\nu_2$, which is too complicated to present here. The
 eigenvalues of both Points (III.2p) and (III.2m) read
 \be{eq66}
 \left\{0,~\frac{3}{4}\left[\,2+2\gamma+\eta_3
 +3(\gamma-2)\sigma_3-\sqrt{\sigma_{32}}\, \right]\,,~
 \frac{3}{4}\left[\,2+2\gamma+\eta_3+3(\gamma-2)\sigma_3
 +\sqrt{\sigma_{32}} \, \right]\right\}\,,
 \ee
 where $\sigma_{32}$ is a function of $\gamma$, $\eta_3$ and
 $\nu_2$, which is too complicated to present here. All these
 four critical points can exist and are stable under some
 conditions, which are very complicated and we do not present
 them here. Nevertheless, we would like to say more. As
 mentioned above, if $\gamma>2/3$, the universe is certainly
 decelerated ($q>0$) at all these four critical points.
 Unfortunately, the universe cannot be accelerated
 if $\gamma=1$ (pressureless matter) and $4/3$ (radiation),
 even when these four critical points can exist and are
 scaling attractors.


 \begin{table}[tbp]
 \begin{center}
 \begin{tabular}{l|c}
 \hline\hline &\\[-4mm]
 ~Label~ & ~~~Critical Point
 $(\bar{x}_1, \bar{x}_2, \bar{x}_3, \bar{x}_4)$~~~\\[0.5mm]
 \hline & \\[-3mm]
 ~IV.1p~ & ~ $\disp\sqrt{\frac{1}{2}\left(1+\sigma_4\right)},
 \ 0,\ 0,\ \sqrt{\frac{1}{2}\left(1-\sigma_4\right)}$ ~ \\[4mm]
 ~IV.1m~ & ~ $\disp-\sqrt{\frac{1}{2}\left(1+\sigma_4\right)},
 \ 0,\ 0,\ \sqrt{\frac{1}{2}\left(1-\sigma_4\right)}$ ~ \\[4mm]
 ~IV.2p~ & ~ $\disp\sqrt{\frac{1}{2}\left(1-\sigma_4\right)},
 \ 0,\ 0,\ \sqrt{\frac{1}{2}\left(1+\sigma_4\right)}$ ~ \\[4mm]
 ~IV.2m~ & ~ $\disp-\sqrt{\frac{1}{2}\left(1-\sigma_4\right)},
 \ 0,\ 0,\ \sqrt{\frac{1}{2}\left(1+\sigma_4\right)}$ ~ \\[3mm]
 \hline\hline
 \end{tabular}
 \end{center}
 \caption{\label{tab6} Critical points for the autonomous
 system (\ref{eq48})---(\ref{eq51}), in the
 Case (IV) $Q=\eta_4\theta\rho_{\rm tot}$.}
 \end{table}



\subsection{Case (IV) $Q=\eta_4\theta\rho_{\rm tot}$}\label{sec4d}

The fourth choice of interaction is $Q=\eta_4\theta\rho_{\rm tot}$.
 In this case, the corresponding
 $Q_1=\frac{9}{4}\nu_2^{-2}\eta_4 x_1^{-1}$
 and $Q_2=\frac{9}{4}\nu_2^{-2}\eta_4 x_4^{-1}$. There are four
 critical points, and we present them in Table~\ref{tab6}.
 Note that we have introduced a new constant
 \be{eq67}
 \sigma_4\equiv\frac{\sqrt{(\gamma-2)\left[
 (\gamma-2)\nu_2^2-6\eta_4\right]\,}}{(\gamma-2)\nu_2}\,.
 \ee
 Points (IV.1p) and (IV.1m) are scaling solutions, and
 their physical quantities read
 \bea
 &\disp\Omega_\phi=1-\frac{\nu_2^2}{3}(1-\sigma_4)\,,~~~~~~~
 \Omega_m=\frac{\nu_2^2}{3}(1-\sigma_4)\,,\nonumber\\[2mm]
 &\disp w_{\rm tot}=\frac{1}{2}\left[\gamma+
 (2-\gamma)\sigma_4\right]\,,~~~~~~~
 w_\phi=\gamma-1+\frac{3(\gamma-2)(1+\sigma_4)}{2\left[
 \nu_2^2(1-\sigma_4)-3\right]}\,,\nonumber\\[1mm]
 &\disp q=\frac{1}{4}\left[2+3\gamma+3\sigma_4(2-\gamma)
 \right]\,.\label{eq68}
 \eea
 Points (IV.2p) and (IV.2m) are also scaling solutions, and
 their physical quantities read
 \bea
 &\disp\Omega_\phi=1-\frac{\nu_2^2}{3}(1+\sigma_4)\,,~~~~~~~
 \Omega_m=\frac{\nu_2^2}{3}(1+\sigma_4)\,,\nonumber\\[2mm]
 &\disp w_{\rm tot}=\frac{1}{2}\left[\gamma+
 (\gamma-2)\sigma_4\right]\,,~~~~~~~
 w_\phi=\gamma-1+\frac{3(\gamma-2)(1-\sigma_4)}{2\left[
 \nu_2^2(1+\sigma_4)-3\right]}\,,\nonumber\\[1mm]
 &\disp q=\frac{1}{4}\left[2+3\gamma+3\sigma_4(\gamma-2)
 \right]\,.\label{eq69}
 \eea
 Since $0<\gamma<2$, all these four critical points can exist
 under the condition $\eta_4\geq 0$ and $0\leq\Omega_m\leq 1$.
 Thus, $-1\leq\sigma_4\leq 0$. From Eq.~(\ref{eq68}), if
 $\gamma>2/3$, the universe is certainly decelerated ($q>0$)
 at Points (IV.1p) and (IV.1m). Unfortunately, the universe
 cannot be accelerated if $\gamma=1$ (pressureless matter)
 and $4/3$ (radiation). From Eq.~(\ref{eq69}), the universe
 is always decelerated ($q>0$) at Points (IV.2p) and (IV.2m),
 since $0<\gamma<2$ and $\sigma_4\leq 0$.

We then consider the stability of these critical points. The
 corresponding
 $\delta Q_1=-\frac{9}{4}\nu_2^{-2}\eta_4 \bar{x}_1^{-2}\delta x_1$
 and
 $\delta Q_2=-\frac{9}{4}\nu_2^{-2}\eta_4\bar{x}_4^{-2}\delta x_4$.
 The eigenvalues of both Points (IV.1p) and (IV.1m) are
 \be{eq70}
 \left\{0,~\frac{3}{4}\left[2+2\gamma-3(\gamma-2)\sigma_4+
 \frac{\sqrt{\sigma_{41}}}{\eta_4(\gamma-2)^2\nu_2^2}\right]\,,
 ~\frac{3}{4}\left[2+2\gamma-3(\gamma-2)\sigma_4
 -\frac{\sqrt{\sigma_{41}}}{\eta_4(\gamma-2)^2\nu_2^2}
 \right]\right\}\,,
 \ee
 where $\sigma_{41}$ is a function of $\gamma$, $\eta_4$ and
 $\nu_2$, which is too complicated to present here. Points
 (IV.1p) and (IV.1m) can exist and are stable under some
 conditions, which are very complicated and we do not present
 them here. Nevertheless, we would like to say more. Since
 $0<\gamma<2$, $\eta_4\geq 0$ and $-1\leq\sigma_4\leq 0$
 if Points (IV.1p) and (IV.1m) can exist, we find that
 $2+2\gamma-3(\gamma-2)\sigma_4\geq 5\gamma-4$. So, if
 $\gamma>4/5$, at least one of the last two eigenvalues
 is positive (when $\sigma_{41}$ is a real number), or the real
 parts of both last two eigenvalues are positive
 (when $\sigma_{41}$ is an imaginary number). Therefore, Points
 (IV.1p) and (IV.1m) are certainly unstable for $\gamma>4/5$.
 So, if $\gamma=1$ (pressureless matter) and $4/3$ (radiation),
 Points (IV.1p) and (IV.1m) cannot be attractors. They can
 be scaling attractors only when $\gamma<4/5$ (necessary
 but not sufficient condition). On the other hand,
 the eigenvalues of both Points (IV.2p) and (IV.2m) are
 \be{eq71}
 \left\{0,~\frac{3}{4}\left[2+2\gamma+3(\gamma-2)\sigma_4+
 \frac{\sqrt{\sigma_{42}}}{\eta_4(\gamma-2)^2\nu_2^2}\right]\,,
 ~\frac{3}{4}\left[2+2\gamma+3(\gamma-2)\sigma_4
 -\frac{\sqrt{\sigma_{42}}}{\eta_4(\gamma-2)^2\nu_2^2}
 \right]\right\}\,,
 \ee
 where $\sigma_{42}$ is a function of $\gamma$, $\eta_4$ and
 $\nu_2$, which is too complicated to present here. Since
 $0<\gamma<2$, $\eta_4\geq 0$ and $-1\leq\sigma_4\leq 0$
 if Points (IV.2p) and (IV.2m) can exist, we find that
 $2+2\gamma+3(\gamma-2)\sigma_4>0$ always. Thus, at least one
 of the last two eigenvalues is positive (when $\sigma_{42}$
 is a real number), or the real parts of both last two
 eigenvalues are positive (when $\sigma_{42}$ is an imaginary
 number). Therefore, Points (IV.2p) and (IV.2m) are always
 unstable if they can exist.


\section{Concluding remarks}\label{sec5}

In this work, we studied the cosmological evolution of the
 {\AE} models with the power-law-like potential given in
 Eq.~(\ref{eq10}). In the case without matter, there are two
 attractors which correspond to an inflationary universe in
 the early epoch, or a de~Sitter universe in the late time.
 In the case with matter but there is no interaction between
 dark energy and matter, there are only two
 de~Sitter attractors, and no scaling attractor exists. So,
 it is difficult to alleviate the cosmological coincidence
 problem. Therefore, we then allow the interaction between
 dark energy and matter. In this case, several scaling
 attractors can exist under some complicated conditions, and
 hence the cosmological coincidence problem could
 be alleviated. However, their stability and/or the condition
 to accelerate the universe usually require $\gamma<2/3$,
 $4/5$ or $1/2$ (depending on the particular form
 of interaction $Q$). So, it is invalid for the normal matter,
 namely $\gamma=1$ (pressureless matter) and $4/3$ (radiation).

Some remarks are in order. First, in this work we only
 considered the the power-law-like potential given in
 Eq.~(\ref{eq10}), namely $r\leq 2$. In fact, it might
 have new interesting results if we use the power-law-like
 potential in Eq.~(\ref{eq9}) with higher $r$. Second, we
 can even use a more general power-law-like potential,
 \be{eq72}
 V(\theta, \phi)=V_0\,\phi^n+\sum\limits_{\alpha,\,\beta}
 \mu_{\alpha\beta} \theta^\alpha \phi^\beta\,.
 \ee
 Third, in addition to the interaction terms considered here,
 there are many exotic interaction terms in the literature,
 such as $Q\propto\theta(\alpha\rho_m+\beta\rho_\phi)$,
 $Q\propto\rho_m^\alpha \rho_\phi^\beta$, and
 $Q\propto q(\alpha\dot{\rho}+\beta\theta\rho)$, where $q$ is
 the deceleration parameter and $\rho$ can be $\rho_m$, $\rho_\phi$
 or $\rho_{\rm tot}$. Fourth, we can also consider a non-flat
 FRW universe with $k\not=0$. All the above generalizations
 might have novel results, and deserve further investigation.
 Finally, it is of interest to study the phase map of the
 autonomous system more carefully. For instance, one could
 try to discuss the sepatrices (if any) and the perturbations
 around them (we thank the anonymous referee for pointing out
 this issue). However, because our main goal of the present
 work is to find whether the scaling attractors exist to
 alleviate the cosmological coincidence problem, this type
 of discussions seems to be beyond our scope, and the relevant
 discussions might greatly extend the length of this paper. We
 consider that it is better to study this issue in the future works.


\section*{ACKNOWLEDGEMENTS}
We thank the anonymous referee for quite useful comments and
 suggestions, which helped us to improve this work. We are
 grateful to Professors Rong-Gen~Cai and Shuang~Nan~Zhang
 for helpful discussions. We also thank Minzi~Feng, as well as
 Long-Fei~Wang, Xiao-Jiao~Guo, Zu-Cheng~Chen and Jing~Liu for
 kind help and discussions. This work was supported in
 part by NSFC under Grants No.~11175016 and No.~10905005,
 as well as NCET under Grant No.~NCET-11-0790, and the
 Fundamental Research Fund of Beijing Institute of Technology.

\renewcommand{\baselinestretch}{1.12}


\end{document}